\documentclass[conference,9pt,letterpaper]{IEEEtran}

\newif\ifanonymous

\usepackage[utf8]{inputenc}
\usepackage[T1]{fontenc}
\usepackage{cite}
\usepackage{amsmath,amssymb,amsfonts}
\usepackage{algorithmic}
\usepackage{algorithm}
\usepackage{graphicx}
\usepackage{textcomp}
\usepackage{xcolor}
\usepackage{booktabs}
\usepackage{array}
\usepackage{makecell}
\usepackage{enumitem}
\usepackage{subcaption}
\usepackage{url}
\usepackage{xspace}
\usepackage{tikz}
\usepackage[table]{xcolor}
\usetikzlibrary{arrows.meta,positioning,calc}
\usepackage[hidelinks]{hyperref}
\graphicspath{{figures_compact/}{figures_final/}{figures_results/}}

\hypersetup{pdfauthor={},pdftitle={},pdfsubject={},pdfkeywords={},pdfcreator={},pdfproducer={}}

\newcommand{\sys}{VeSGA\xspace}
\newcommand{\vsg}{VS-Graph\xspace}

\definecolor{cCompute}{HTML}{DCE8F5}
\definecolor{cComputeLine}{HTML}{2F5F8F}
\definecolor{cCtrl}{HTML}{F0F0F0}
\definecolor{cCtrlLine}{HTML}{6E6E6E}
\colorlet{perfVeSGA}{cCompute!55!white}
\newcommand{\vesgacell}[1]{\cellcolor{perfVeSGA}\textbf{#1}}

\tikzset{
  blk/.style={draw=cComputeLine, fill=cCompute, line width=0.5pt, rounded corners=1pt,
              align=center, font=\footnotesize, inner sep=2.5pt, minimum height=5mm},
  ctl/.style={blk, draw=cCtrlLine, fill=cCtrl},
  flow/.style={-{Stealth[length=1.6mm,width=1.3mm]}, line width=0.6pt},
  num/.style={circle, fill=cComputeLine, text=white, font=\bfseries\footnotesize,
              inner sep=0.5pt, minimum size=3.4mm},
  lbl/.style={font=\footnotesize, inner sep=1pt},
}
\newcommand{\circled}[1]{\tikz[baseline=(c.base)]\node[num,minimum size=2.8mm,font=\bfseries\scriptsize](c){#1};}

\begin{document}

\title{VeSGA: Microsecond, Microjoule Hyperdimensional Graph Classification through Algorithm-Hardware Co-Design}

\ifanonymous
  \author{\IEEEauthorblockN{Anonymous Authors}
         \IEEEauthorblockA{Paper under double-blind review}}
\else
  
  \author{
    \IEEEauthorblockN{
        Ahmad Tahmasivand\textsuperscript{1} \quad
        Hamed Poursiami\textsuperscript{1} \quad
        Guojing Cong\textsuperscript{2} \quad
        Catherine Schuman\textsuperscript{2} \\
        Gina Adam\textsuperscript{3} \quad
        Thomas Potok\textsuperscript{2} \quad
        Khaled N. Khasawneh\textsuperscript{1} \quad
        Maryam Parsa\textsuperscript{1}
        }
        \IEEEauthorblockA{
        \textsuperscript{1}Electrical and Computer Engineering Department,
        George Mason University, Fairfax, VA, USA \\
        \textsuperscript{2}Oak Ridge National Laboratory, Oak Ridge, TN, USA\\
        \textsuperscript{3}Electrical and Computer Engineering Department,
        George Washington University, Washington, DC, USA
        }
        \IEEEauthorblockA{
        \texttt{atahmasi@gmu.edu, hpoursia@gmu.edu, 
        congg@ornl.gov, cschuman@utk.edu,} \\
        \texttt{ginaadam@gwu.edu, potokte@ornl.gov,
        kkhasawn@gmu.edu, mparsa@gmu.edu}
        }
        
    }
\fi

\maketitle

\begin{abstract}
Deploying graph classification at the edge requires pairing high predictive accuracy with ultra-low latency and energy consumption. While message-passing hyperdimensional computing (HDC) closes the accuracy gap with graph neural networks (GNNs), existing formulations remain designed for software and map poorly onto resource-constrained hardware. We present \sys, an FPGA accelerator co-designed with its algorithm that adapts every stage of HDC graph classification to hardware. It is implemented on the ZCU104 FPGA board and evaluated across six benchmarks. Including data transfer, it runs up to 25$\times$ faster than GPU computation alone, achieves up to 42$\times$ lower latency than the closest prior accelerator, and consumes up to 3\,900$\times$ less dynamic energy than the Nvidia RTX3090 GPU.

\end{abstract}


\section{Introduction}
\label{sec:intro}

Graph classification increasingly needs to run where graphs are produced; on embedded devices with power budgets of a few watts~\cite{graphhd2022}. Graph neural networks (GNNs)~\cite{gilmer2017mpnn,kipf2017gcn,velickovic2018gat,xu2019gin} are the standard tool, but fit this setting poorly~\cite{graphhd2022}. They require gradient-based training and irregular floating-point message passing. On small classification graphs (15 to 300 nodes), their per-graph CPU or GPU overheads exceed the useful work.

Hyperdimensional computing (HDC) is a brain-inspired alternative~\cite{kanerva2009,kleyko2022vsa}. It computes with simple, highly parallel operations on wide binary vectors and learns without gradients, making it well-suited to small devices. Early HDC graph classifiers~\cite{graphhd2022,nyshd2025} trailed GNNs in accuracy, and \vsg~\cite{vsgraph2025} closed most of that gap with message passing in the hyperdimensional space. The published algorithm, however, runs slowly even on a GPU and lags behind competing GNNs on larger graphs~\cite{vsgraph2025}. This is because bit-level operations on wide vectors over tiny irregular graphs map poorly onto floating-point datapaths. Existing HDC graph accelerators target other formulations (Sec.~\ref{sec:background}).

We address this gap by designing the algorithm and hardware together. We reduce the algorithm to simple, error-tolerant bit-level operations and build an accelerator that processes each graph on chip with hundreds of parallel lanes. We present \sys (Vector-Symbolic Graph Classification Accelerator) on an embedded FPGA. Our four contributions are:

\begin{figure}[t]
  \centering
  \includegraphics[width=\columnwidth]{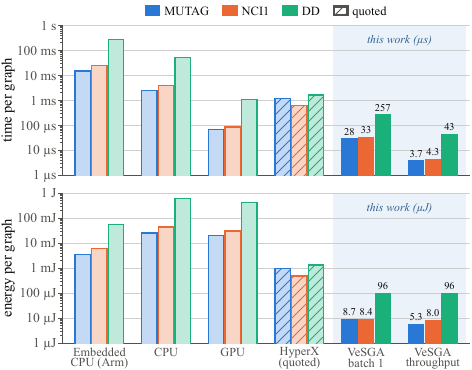}
  \caption{Time and dynamic energy per graph (log scale) on MUTAG, NCI1 and DD. HyperX is quoted~\cite{hyperx2026} (hatched), and \sys runs at batch one and at full throughput.}
  \label{fig:teaser}
\end{figure}

\begin{itemize}[leftmargin=*,itemsep=1pt,topsep=2pt]
  \item \textbf{Algorithm--hardware co-design of HDC message passing} (Sec.~\ref{sec:codesign}). Binary majority-vote aggregation, a hashed rank code book and an integer readout reduce lane width from 16 bits to one and eliminate a 6\,MB table, all multipliers and dividers, while accuracy remains within the fold-to-fold spread on all six benchmarks.
  \item \textbf{An on-chip architecture for HDC message passing} (Sec.~\ref{sec:arch}). Each processing element (PE) keeps a whole graph on chip and streams its Compressed Sparse Row (CSR) edge list at two edges per clock to a 512-lane majority array. A slice schedule writes the next slice's code for free, and exact halo-based splitting extends support beyond the PE capacity.
  \item \textbf{Measured evaluation on an embedded FPGA} (Secs.~\ref{sec:method}, \ref{sec:eval}). On a ZCU104 board, \sys classifies a graph in tens of microseconds using microjoules of energy, matching the software reference bit for bit. It is up to 25$\times$ faster than the same kernel on a GPU, up to 750$\times$ more energy-efficient than the software on the board's embedded CPU, and up to 42$\times$ lower in latency than the closest prior accelerator (Fig.~\ref{fig:teaser}, Table~\ref{tab:perf}).
  \item \textbf{The first characterization of an HDC graph accelerator beyond nominal operating conditions} (Sec.~\ref{sec:eval:robust}). Undervolting by 23\,\% is bit-exact and halves the dynamic energy. The first timing failure is confined to one lane group in one PE without accuracy loss, and the datapath degrades gracefully until the control path hangs.
\end{itemize}

\section{Background and Related Work}
\label{sec:background}
\label{sec:related}

\textbf{Hyperdimensional graph classification.} HDC is a brain-inspired paradigm that distributes information across high-dimensional vectors called hypervectors. Binary HDC bundles hypervectors by a dimension-wise majority vote, binds them with XOR and compares them by Hamming or cosine similarity. These simple, parallel operations make HDC efficient in hardware, and the distributed representation tolerates errors in individual components~\cite{kanerva2009,thomas2021theory}. GraphHD~\cite{graphhd2022} assigns node hypervectors by PageRank~\cite{brin1998pagerank} rank and bundles the bound hypervectors of connected nodes. NysHD~\cite{nyshd2025} approximates a graph kernel by a Nystr\"om projection, and HDGL~\cite{hdgl2025} and HyperGraphX~\cite{hypergraphx2025} classify nodes. \vsg~\cite{vsgraph2025} adds iterative neighborhood aggregation. Its \emph{spike-diffusion} stage scores the nodes over $K$ hops, and the rank of each node selects its initial binary hypervector from a shared code book ($D=8192$ here). In each of $L$ \emph{associative message-passing} layers, every node blends the aggregated hypervectors of its neighbors with its own state, and the mean over all nodes forms the graph embedding. A graph is assigned to the class whose mean training embedding is most similar in cosine, so learning needs no gradients. In topology-only evaluation, \vsg is competitive with or better than GCN~\cite{kipf2017gcn}, GAT~\cite{velickovic2018gat} and GIN~\cite{xu2019gin} on the TU datasets~\cite{tudataset2020} and keeps its accuracy down to $D=128$~\cite{vsgraph2025}.

\textbf{Where the time goes.} The published implementation (PyTorch~\cite{paszke2019pytorch} and DGL~\cite{wang2019dgl} on a Tesla T4) takes 0.33 to 0.42\,ms per graph on the small benchmarks and 2.3\,ms on DD, $4$ to $7\times$ slower than GNNs on DD~\cite{vsgraph2025}. We attribute this to kernel-launch overhead on 18-node graphs, not to the arithmetic. CPUs fare no better, as the NumPy~\cite{harris2020numpy} reference takes 15\,ms per MUTAG graph on an embedded CPU and 2.4\,ms on an Intel Core i5 CPU. Message passing accounts for nearly all of the roughly 50 passes a graph makes over its edge list (Sec.~\ref{sec:arch:slices}), so an accelerator must speed it up without letting diffusion, sorting and readout dominate small graphs.

\textbf{HDC hardware for graphs.} HyperX~\cite{hyperx2026} runs NysHD on the same ZCU104 and streams a floating-point projection matrix from DDR. RelHD~\cite{relhd2022} and HDGAS~\cite{hdgas2026} target node-level learning in simulated processing-in-memory and ASIC designs, HyperGRAF~\cite{hypergraf2023} and HDReason~\cite{hdreason2024} graph reasoning on data-center FPGAs, and HygHD~\cite{hyghd2024} hypergraphs. FPGA undervolting has been studied for neural-network accelerators~\cite{salami2020undervolt}, and HDC is expected to tolerate the resulting errors~\cite{kleyko2022vsa}. To our knowledge, \sys is the first accelerator for message-passing HDC graph classification and the first HDC graph accelerator characterized under undervolting and overclocking.

\textbf{Why not a GNN accelerator.} GNN accelerators build on weight matrices and multiply-accumulate arrays, and HyGCN~\cite{yan2020hygcn}, AWB-GCN~\cite{geng2020awbgcn} and I-GCN~\cite{geng2021igcn} classify the nodes of one large graph with thousands of multipliers. FlowGNN~\cite{flowgnn2023} streams small molecular graphs on a data-center card at 20 to 60\,$\mu$s per graph, and the only embedded GNN result on the TU datasets we know of takes 8 to 14\,ms~\cite{grailoo2025}. The \vsg kernel has no weights or multiplications. Its aggregation is a count, its update a compare and its code book a hash. It needs one bit of on-chip state per node and lane, two random reads per clock into that state and hundreds of identical one-bit lanes, which GNN templates are not built to provide. As these accelerators run different models, we claim no speed-up over them.

\section{Algorithm--Hardware Co-Design}
\label{sec:codesign}

\begin{figure}[t]
  \centering
  \begin{tikzpicture}[x=1cm,y=1cm,font=\footnotesize,
  stage/.style={ctl, text width=1cm, minimum height=0.75cm, align=center},
  pub/.style={draw=cCtrlLine, fill=white, line width=0.5pt, rounded corners=1pt,
              text width=3.05cm, minimum height=0.75cm, align=left, inner sep=2.5pt},
  ours/.style={blk, text width=3.15cm, minimum height=0.75cm, align=left}]
\node[lbl,font=\footnotesize\bfseries] at (-0.4,0.7) {Stage};
\node[lbl,font=\footnotesize\bfseries] at (2.18,0.7) {VS-Graph (published)};
\node[lbl,font=\footnotesize\bfseries,text=cComputeLine] at (5.82,0.7) {VeSGA kernel};
\node[stage] (s0) at (-0.4,0)    {spike\\diffusion};
\node[stage] (s1) at (-0.4,-1.05) {rank};
\node[stage] (s2) at (-0.4,-2.25) {rank code};
\node[stage] (s3) at (-0.4,-3.6) {message\\passing $\times L$};
\node[stage] (s4) at (-0.4,-4.8) {readout};
\node[pub] (p0) at (2.18,0)    {$\mathbf{s}\leftarrow A^{K}\mathbf{1}$\\real-valued scores};
\node[pub] (p1) at (2.18,-1.05) {sort nodes by score};
\node[pub] (p2) at (2.18,-2.25) {$\mathbf{h}^{(0)}_v = B[r_v]$\\stored table\\$n_{\max}\times D$ bits};
\node[pub] (p3) at (2.18,-3.6) {$m_v=\bigvee_{u\in\mathcal N(v)} h_u$\\$h_v\leftarrow\alpha h_v+(1\!-\!\alpha)\,m_v$\\real-valued state};
\node[pub] (p4) at (2.18,-4.8) {$\mathbf{z}=\tfrac{1}{n}\sum_v \mathbf{h}_v$\\floating-point mean};
\node[ours] (o0) at (5.82,0)    {uint16 scores\\hop 1 = in-degree};
\node[ours] (o1) at (5.82,-1.05) {stable LSD radix sort\\$\mathcal{O}(n)$, two 8-bit digits};
\node[ours,line width=1pt] (o2) at (5.82,-2.25) {$B[r]_w=\mathrm{hash}(\sigma,r,w)$\\generated on the fly\\0 bytes stored};
\node[ours,line width=1pt] (o3) at (5.82,-3.6) {majority vote per lane\\$\mathrm{MAJ}\big(h^{(0)}_v,\{h_u\}_{u\in\mathcal N(v)}\big)$\\1 bit per lane, ties $\to 1$};
\node[ours] (o4) at (5.82,-4.8) {$z[d]=2\,\mathrm{cnt}[d]-n$\\shift by $\lfloor\log_2 n\rfloor$, int16};
\foreach \i/\j in {0/1,1/2,2/3,3/4} {
  \draw[flow] (s\i.south) -- (s\j.north);
  \draw[flow] (p\i.south) -- (p\j.north);
  \draw[flow] (o\i.south) -- (o\j.north);
}
\foreach \i/\n in {0/A,1/B,2/C,3/D,4/E} {
  \node[num] at ([xshift=-1.2mm,yshift=-0.4mm]o\i.north east) {\n};
}
\node[ctl, text width=6.9cm, minimum height=0.5cm, align=center] (cls) at (3.9,-5.8)
  {class-mean prototypes, cosine similarity (host, unchanged)};
\draw[flow] (p4.south) -- (p4.south |- cls.north);
\draw[flow] (o4.south) -- (o4.south |- cls.north);
\end{tikzpicture}
  \caption{The \vsg pipeline (left), the published formulation~\cite{vsgraph2025} (center) and the co-designed \sys kernel (right). The two changes that shape the architecture, hashed codes \protect\circled{C} and the binary majority vote \protect\circled{D}, are outlined.}
  \label{fig:codesign}
\end{figure}
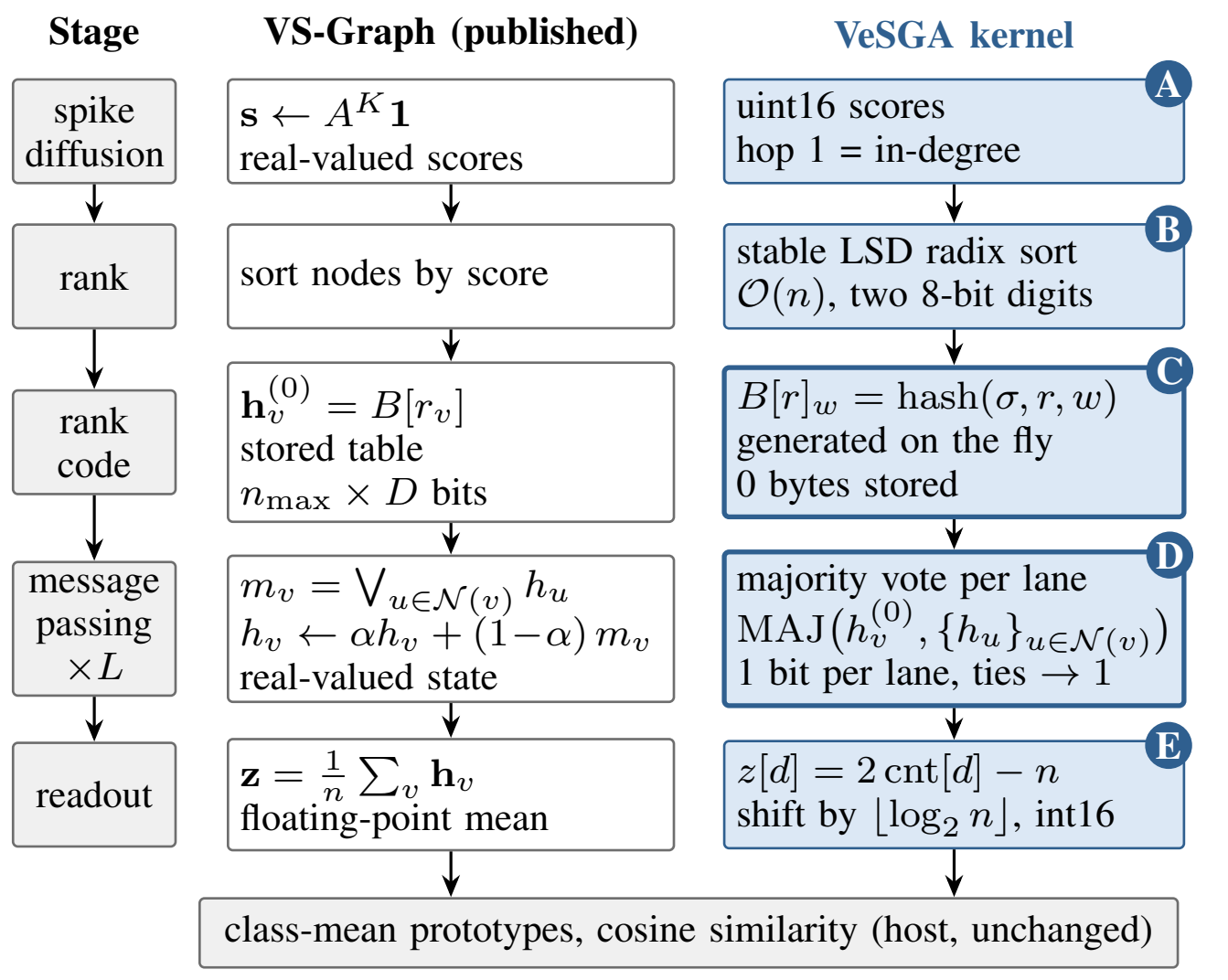

Fig.~\ref{fig:codesign} summarizes the \vsg pipeline and our changes. A graph has $n$ nodes ($n_{\max}$ for the largest one), and $h^{(l)}_v[d]$ is dimension $d=0,\dots,D-1$ of the hypervector of node $v$ after layer $l$, where each dimension is a lane in hardware. The published formulation is a poor hardware target. It keeps a real-valued state per node and lane, stores a code book of $n_{\max}\times D$ random bits, and divides by $n$ in floating point in the readout. We co-designed the kernel and accelerator around one rule: \emph{every node state is one bit per lane, every constant derives from the rank, and every operation is an add or a compare}. Five changes (\circled{A}--\circled{E} in Fig.~\ref{fig:codesign}) implement this rule (Algorithm~\ref{alg:kernel}).

\subsection{Binary Majority-Vote Message Passing}
\label{sec:codesign:maj}
The published update (Fig.~\ref{fig:codesign}, center) aggregates the neighbors with a dimension-wise OR into $\mathbf{m}_v$ and blends it with the previous state, $\mathbf{h}_v\leftarrow\alpha\mathbf{h}_v+(1-\alpha)\mathbf{m}_v$, where the blend weight $\alpha\in[0,1]$ controls how much of its previous state a node retains. The state becomes real-valued after the first layer, so hardware carries $b$ bits per node and lane ($b\ge L+1$ even for $\alpha=\tfrac12$, 8 to 16 bits in general fixed point), a $b$-bit maximum per edge and a blend per node. \sys replaces aggregation and blend with a binary majority vote (\circled{D}) in which the node's own rank code acts as the residual,
\begin{equation}
h^{(l+1)}_v[d]=\Big[\,h^{(0)}_v[d]+\!\!\sum_{u\in\mathcal N^-(v)}\!\! h^{(l)}_u[d]\;\ge\;\Big\lceil\tfrac{\deg^-(v)+1}{2}\Big\rceil\Big],
\label{eq:maj}
\end{equation}
where $[\cdot]$ is the Iverson bracket, $\mathcal N^-(v)$ are the in-neighbors of $v$, $\deg^-(v)=|\mathcal N^-(v)|$ is its in-degree, and ties resolve to one. Eq.~\eqref{eq:maj} is HDC bundling~\cite{kanerva2009} of the multiset $\{h^{(0)}_v\}\cup\{h^{(l)}_u\}_{u\in\mathcal N^-(v)}$. The self vote anchors the node to its structural identity, as $\alpha$ does, and the degree-dependent threshold normalizes the aggregate, as the convex weights $\alpha$ and $1-\alpha$ do. Because bundling is idempotent under repeated identical votes, the state neither saturates nor drifts across layers. Per edge and lane, the hardware adds one bit into a $\lceil\log_2(\deg^-_{\max}+2)\rceil$-bit counter, where $\deg^-_{\max}$ is the maximum in-degree (five bits for the supported maximum of 30, which no benchmark graph exceeds), and per node and lane it performs one comparison. Lanes never interact, so $D$ splits into independent slices of $W$ lanes (Sec.~\ref{sec:arch:slices}) and the state of a 4096-node graph stays on chip.

\subsection{Hashed Rank Code Book}
\label{sec:codesign:hash}
\vsg draws the code of rank $r$ from a pseudo-random table $B\in\{0,1\}^{n_{\max}\times D}$, 5.9\,MB for the largest DD graph at $D=8192$, and gathers one row per node. \sys stores no table; it computes each code on the fly (\circled{C}), with word $w$ of the code of rank $r$ given by
\begin{equation}
B[r]_w=\mathrm{lowbias32}\big(\sigma\oplus(r\ll 16)\oplus w\big),\qquad w=0,\dots,D/32-1,
\label{eq:hash}
\end{equation}
where $\sigma$ is a 16-bit seed from the record header and lowbias32 is a 32-bit integer mixer~\cite{wellons2018hash} of two constant multiplies and three xor-shifts, mapped onto pipelined DSP slices. $W/32$ hash units produce a $W$-lane code slice at one node per clock in lockstep with the CSR walk, so codes are never fetched from memory. As a pure function of $(\sigma,r,w)$, the code is identical on the host and FPGA, independent of the dataset and $n$, and defined for every rank up to $2^{16}$. A part of a large graph can therefore recreate the exact codes of the whole graph from its global ranks (Sec.~\ref{sec:arch:parts}).

\subsection{Integer Diffusion, Ranking and Readout}
\label{sec:codesign:int}
Diffusion scores (\circled{A}) are 16-bit unsigned integers with wrap-around. The first hop uses the in-degree, and each further hop sums the in-neighbors' scores, two per clock, using the walker that drives message passing. A two-digit least-significant-digit (LSD) radix sort (\circled{B}) stably ranks the scores in descending order in about $9n$ cycles and breaks ties by node order, as the software reference does. The readout (\circled{E}) counts the ones in each lane in 13 bits and emits the embedding $z[d]=2\,\mathrm{cnt}[d]-n$ scaled by $2^{8}/2^{\lfloor\log_2 n\rfloor}$, a shift instead of a division. This scale is within a factor of two of $1/n$. The classifier is invariant to a common scale, and the shift changes accuracy by at most 0.5 points (Sec.~\ref{sec:eval:acc}).

\begin{algorithm}[t]
\caption{\sys kernel for one graph (one record)}
\label{alg:kernel}
\begin{algorithmic}[1]
\REQUIRE $n$, in-edge CSR (row pointers $\mathrm{indptr}$, source ids $\mathrm{src}$), hops $K$, layers $L$, seed $\sigma$; optional: global ranks $r$, core count $n_\mathrm{core}$ (Sec.~\ref{sec:arch:parts})
\IF{ranks not supplied}
  \STATE diffusion score $s_v\leftarrow\deg^-(v)$;\quad \textbf{for} $t=2..K$: $s_v\leftarrow\sum_{u\in\mathcal N^-(v)} s_u$ \hfill (uint16)
  \STATE $r\leftarrow$ stable rank of $s$, descending \hfill (radix sort)
\ENDIF
\FOR{slice $j=0..D/W-1$}
  \STATE $h^{(0)}_v\leftarrow B[r_v]_{\,jW/32\,..\,(j+1)W/32-1}$ for all $v$ \hfill (Eq.~\ref{eq:hash}, no table)
  \FOR{layer $l=0..L-1$}
    \STATE $h^{(l+1)}_v\leftarrow\mathrm{MAJ}\big(h^{(0)}_v,\{h^{(l)}_u\}_{u\in\mathcal N^-(v)}\big)$ for all $v$ \hfill (Eq.~\ref{eq:maj})
  \ENDFOR
  \STATE $\mathrm{cnt}[jW{+}d]\leftarrow\sum_{v<n_\mathrm{core}} h^{(L)}_v[d]$ \hfill (13-bit counters)
\ENDFOR
\RETURN $(2\,\mathrm{cnt}-n)\ll 8\gg\lfloor\log_2 n\rfloor$ \hfill (or $2\,\mathrm{cnt}$ raw for a part)
\end{algorithmic}
\end{algorithm}

\subsection{What the Changes Remove and What They Cost}
\label{sec:codesign:cost}
At the design capacity of 4096 nodes per PE, the binary state shrinks the ping-pong hypervector buffers of a PE from 32--64\,Mbit to 4\,Mbit in 16 UltraRAMs. With equal on-chip memory, the published formulation would process only 32 to 64 lanes per clock instead of 512 and would need a multiplier per lane unless $\alpha=\tfrac12$. The hashed code book removes the only large table and its per-node gather, and the integer readout removes the only divider. Accuracy remains comparable (Table~\ref{tab:codesign_acc}, six TU benchmarks~\cite{tudataset2020}, identical folds). The FPGA kernel with hashed codes (C) lies within the fold-to-fold spread of the same kernel with PRNG codes (B) and of the published formulation (A) on every dataset (Sec.~\ref{sec:eval:acc}).

\begin{table}[t]
\caption{Accuracy (\%, mean $\pm$ std over 30 folds) of the published formulation (A) and the hardware kernel with PRNG codes (B) or hashed codes on the FPGA (C).}
\label{tab:codesign_acc}
\centering
\scriptsize
\setlength{\tabcolsep}{2pt}
\resizebox{\columnwidth}{!}{%
\begin{tabular}{@{}l@{\hspace{3pt}}cccccc@{}}
\toprule
Variant & MUTAG & PTC\_FM & ENZYMES & PROTEINS & NCI1 & DD \\
\midrule
published$^\dagger$ & 88.5 & 60.4 & -- & 73.3 & 63.2 & 76.5 \\
A: sum $+$ blend & $83.9{\pm}8.9$ & $59.6{\pm}7.3$ & $21.9{\pm}5.1$ & $71.6{\pm}3.7$ & $61.5{\pm}2.1$ & $75.2{\pm}4.4$ \\
B: PRNG codes & $85.9{\pm}8.9$ & $60.4{\pm}6.4$ & $23.4{\pm}4.9$ & $72.0{\pm}3.5$ & $63.1{\pm}2.1$ & $75.5{\pm}4.2$ \\
\textbf{C: hashed (FPGA)} & $85.1{\pm}8.8$ & $60.2{\pm}6.2$ & $23.4{\pm}4.7$ & $71.8{\pm}3.6$ & $63.2{\pm}2.2$ & $75.5{\pm}4.2$ \\
\bottomrule
\end{tabular}}
\par\smallskip
\raggedright\scriptsize $^\dagger$As reported in~\cite{vsgraph2025} (own folds; ENZYMES not reported).
\end{table}
\section{\sys Architecture}
\label{sec:arch}

\begin{figure*}[t]
  \centering
  \includegraphics[width=0.9\textwidth]{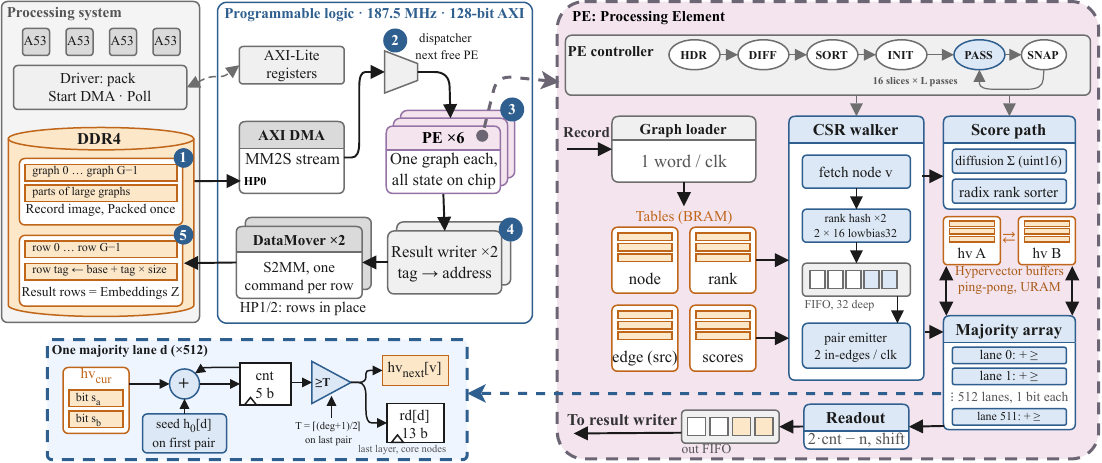}
  \caption{\sys system overview (left), one processing element (right) and one of its 512 majority lanes (bottom left). Gray marks the processing system and vendor IP, blue our logic and orange memory.}
  \label{fig:system}
\end{figure*}

\sys maps Algorithm~\ref{alg:kernel} onto an accelerator that keeps each graph on chip (Fig.~\ref{fig:system}). The programmable logic contains an input DMA, a dispatcher, six processing elements (PEs), and two result writers. A small register block stores configuration and status. Each record header carries the kernel parameters ($K$, $L$, $\sigma$), allowing a batch to mix datasets, whole graphs and parts of large graphs.

\subsection{System and Data Flow}
\label{sec:arch:system}
The host packs a dataset once into an image of \emph{records}, each containing a header and the in-edge list of one graph in CSR form (\circled{1}). One DMA transfer streams the image to the dispatcher, which sends each record to the next free PE (\circled{2}). A PE embeds one graph at a time with all its state on chip, so a large graph never stalls the other PEs (\circled{3}). Results leave the PEs out of order, tagged with their record indices. The result writers place each at the corresponding memory row (\circled{4}). Memory thus restores dataset order. For each batch, the host starts one transfer, polls two row counters and synchronizes the cache (\circled{5}).

\subsection{Processing Element}
\label{sec:arch:pe}
A PE holds a graph of up to 4096 nodes and 16\,384 edges (Fig.~\ref{fig:system}, right). The \emph{graph loader} writes the record into block-RAM tables for nodes, edges, ranks and scores, and two UltraRAM buffers hold the hypervectors of the current and the next layer. A controller steps through the stages of Algorithm~\ref{alg:kernel}, from the header (HDR) through diffusion (DIFF), ranking (SORT) and the first codes (INIT) to the message-passing passes (PASS) and the output of each slice (SNAP).

The \emph{CSR walker} is the only loop in hardware. For each node it reads the rank, hashes it into the node's code (Eq.~\eqref{eq:hash}) and then emits two of the node's in-edges per clock. The other units consume this stream without waiting for a code or an address. The \emph{score path} sums the diffusion scores and ranks them with a radix sort. This stage is skipped when a record supplies its ranks. The \emph{majority array} has 512 independent lanes (Fig.~\ref{fig:system}, bottom left). For each edge pair, lane $d$ adds the bits of the two source nodes $s_a$ and $s_b$ to a small counter that starts with the node's own code bit, and after the last pair it compares the count with the threshold of Eq.~\eqref{eq:maj} and writes one result bit. Because lanes are independent, the array is simply 512 copies of a small adder and comparator. In the last layer, the result bits are also counted per lane over the graph's nodes (only the core nodes for a part, Sec.~\ref{sec:arch:parts}), and the \emph{readout} turns these counts into the embedding. Two edges per clock nearly double the speed of a one-edge walker at the median in-degrees of two to five, while four would halve the number of PEs that fit within the same UltraRAM budget.

\subsection{Slice Schedule and the Free Next-Slice Code}
\label{sec:arch:slices}
\begin{figure}[t]
  \centering
  \includegraphics[width=0.9\columnwidth]{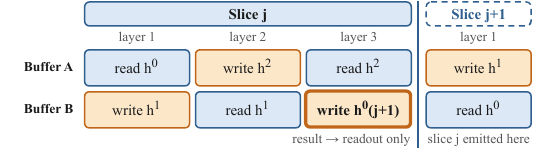}
  \caption{Roles of the two hypervector buffers over two consecutive slices ($L=3$; blue read, orange write).}
  \label{fig:schedule}
\end{figure}
Because lanes are independent, \sys processes the $D=8192$ dimensions as 16 slices of $W=512$ lanes, so a PE needs only two 512-bit-wide buffers instead of one 8192-bit-wide buffer. Within a slice, the $L$ layers alternate the roles of the two buffers (Fig.~\ref{fig:schedule}). Loading the codes of the next slice would normally need a third buffer or an extra pass. However, the last-layer results go only to the readout counters, so \sys writes the codes of the next slice, $h^{(0)}(j{+}1)$, into the free buffer (bold in Fig.~\ref{fig:schedule}), and the next slice starts at once while the current one is emitted. A graph thus needs $1+16L$ passes instead of $16(L{+}1)$, and a PE needs one-third less UltraRAM, which lets six PEs fit on the device.

\subsection{Exact Processing of Graphs Beyond the PE Capacity}
\label{sec:arch:parts}
A graph with more than 4096 nodes or 16\,384 edges does not fit a PE, so \sys splits it into exact parts. This works because Eq.~\eqref{eq:maj} is local, as $h^{(L)}_v$ depends only on the nodes within $L$ hops upstream of $v$, the readout is a sum over nodes, and the codes of Eq.~\eqref{eq:hash} depend only on the global rank. The host ranks the whole graph once and cuts it into \emph{cores}, each extended by its $L$-hop in-neighborhood, the \emph{halo}, so that each part fits a PE. Each part runs as an ordinary record that carries the global ranks and counts only its $n_\mathrm{core}$ core nodes. The host adds the counts of all parts and applies the final scaling, which reconstructs the exact whole-graph embedding (Sec.~\ref{sec:eval:scale}).
\section{Experimental Methodology}
\label{sec:method}

\textbf{Platform.} \sys is implemented in VHDL and built using Vivado 2024.1 for the XCZU7EV-2 on a ZCU104 board~\cite{zcu104ug} (16\,nm Zynq UltraScale+). The host is the on-board quad Cortex-A53 at 1.2\,GHz with Ubuntu 22.04 and a Python driver. No PC is involved in any \sys measurement. All results use six PEs in one 187.5\,MHz clock domain with 128-bit AXI paths, $W=512$, $D=8192$ and a capacity of 4096 nodes and 16\,384 edges per PE. The measurement bitstream adds a cycle counter and four timestamp registers to the production design without other design changes. Before each run, we verify it bit-exactly against the host model on every dataset.

\textbf{Datasets and protocol.} We use the six TU benchmarks~\cite{tudataset2020} of GraphHD~\cite{graphhd2022} and \vsg~\cite{vsgraph2025} in topology-only form (graph counts in Table~\ref{tab:perf}) with the reference settings, $D=8192$, $K=4$ hops, $L=3$ layers and class-mean cosine prototypes. Accuracies are the mean $\pm$ standard deviation over stratified 10-fold cross-validation with three seeds, which also select the code book (30 folds). We process all of DD, splitting its two graphs above the PE capacity into four exact parts (Sec.~\ref{sec:arch:parts}).

\textbf{Timing.} We distinguish two timing quantities. \emph{Batch-1 latency} runs from the first input beat of a single-graph record to its last result row. Cycle counters in the fabric record it for all 7540 records, and we report its mean and 99th percentile (p99). \emph{Throughput} counts graphs per second when a whole dataset streams as one batch with all six PEs busy, timed in hardware (DMA start to last result row, including the transfer of graphs and results) and end to end (Python call). The host overhead of about 27\,$\mu$s per DMA transaction, independent of the batch size, is reported separately.

\textbf{Energy.} We sample the 12\,V board input (INA226) and the 0.85\,V VCCINT core rail (IRPS5401) at 25\,Hz over 4\,s of back-to-back runs. The core rail feeds the programmable logic and the full-power processing domain. Dynamic energy per graph is the power above the idle floor times the time per graph. \emph{Rail dynamic} uses the core rail and isolates the fabric, while \emph{board dynamic} uses the 12\,V input and also covers the processing system, DRAM and regulators. With the bitstream loaded, the idle floor is 12.7\,W on the board and 4.05\,W on the rail. For HyperX we also use its basis, the \emph{tool estimate}, which multiplies the Vivado post-route power of the accelerator by the measured time.

\textbf{Baselines.} All software baselines run the kernel in Python. The CPUs run the NumPy reference with PRNG codes (B in Table~\ref{tab:codesign_acc}) on one pinned core. The \emph{embedded CPU} is an A53 of the same board, measured with the same sensors, and the \emph{CPU} is an Intel Core i5-1135G7, measured by the package energy of the running average power limit (RAPL) interface~\cite{david2010rapl} minus idle. The \emph{GPU} runs a bit-exact PyTorch port of the hardware kernel (C) on an RTX 3090 in batches of up to 200\,k nodes. Its device time excludes the host-to-device transfer and its end-to-end time includes it. Its energy is the board counter of the NVIDIA management library (NVML)~\cite{nvml} minus idle, read over at least 10\,s because the counter updates every 100\,ms. All times are per graph over a whole dataset, averaged over three passes after a warm-up, without the two oversize DD graphs. We quote HyperX~\cite{hyperx2026}, a Nystr\"om-HDC classifier on the same board (Vitis cycle-accurate latency, post-implementation power), and compare it in speed and energy only, since it runs a different algorithm.
\section{Evaluation}
\label{sec:eval}

\subsection{Accuracy}
\label{sec:eval:acc}
FPGA embeddings are bit-identical to the host model on every graph, so their accuracies agree in every fold. Table~\ref{tab:codesign_acc} compares the variants under one protocol. Our run of the published formulation (A) follows its reference implementation (sum aggregation, $\alpha=0.1$, float mean readout), because the paper's OR aggregation causes the blended state to saturate on dense DD graphs and collapse to the majority class. The hardware kernel (C) stays within the fold-to-fold spread of A and of the PRNG-code reference (B) on every dataset. The published numbers use other folds and seeds, which explains their offset from A, and small differences among A, B and C reflect the hardware-oriented approximations of Sec.~\ref{sec:codesign}. The shift readout and the integer diffusion each cost at most 0.5 points, whereas removing the self vote costs up to 2.0 points, so the self vote is the essential part of the blend. The hardware kernel thus keeps the accuracy profile of the published \vsg against topology-only GNNs, including its deficit to GIN on NCI1~\cite{vsgraph2025}.

\subsection{Resources and Timing}
\label{sec:eval:res}
Six PEs use all 96 UltraRAMs and about two thirds of the LUTs, block RAM and DSP slices (Table~\ref{tab:resources}) and meet timing at 187.5\,MHz. The worst paths are dominated by routing and occur in the rank-to-hash and readout nets of one PE, which matters in Sec.~\ref{sec:eval:robust}.

\begin{table}[t]
\caption{Post-place resource use on the XCZU7EV. BRAM in 36\,kb tiles; per PE $=$ accelerator block$/6$.}
\label{tab:resources}
\centering
\footnotesize
\setlength{\tabcolsep}{4pt}
\begin{tabular}{@{}lrrrrr@{}}
\toprule
 & LUT & FF & BRAM & URAM & DSP \\
\midrule
available            & 230\,400 & 460\,800 & 312 & 96 & 1\,728 \\
whole design, 6 PEs  & 143\,954 & 159\,383 & 216 & 96 & 1\,152 \\
\quad share of device & 62\,\% & 35\,\% & 69\,\% & 100\,\% & 67\,\% \\
accelerator block    & 137\,801 & 151\,464 & 192 & 96 & 1\,152 \\
per PE (derived)     & 23.0\,k & 25.2\,k & 32 & 16 & 192 \\
\bottomrule
\end{tabular}
\end{table}

\subsection{Latency and Throughput}
\label{sec:eval:lat}
Table~\ref{tab:perf} lists the time per graph on all platforms. At batch one, \sys embeds a graph of the small benchmarks in tens of microseconds and a DD graph in 257\,$\mu$s, with p99 latencies of at most 155\,$\mu$s on the small benchmarks and 0.9\,ms on DD. At full load, the six PEs execute in parallel, which cuts the hardware time per graph by another $6$ to $8\times$ and sustains up to 341\,k graphs per second. This is at least $440\times$ faster than the software reference on both CPUs. The GPU amortizes kernel launches over thousands of graphs, yet its device time alone is $11$ to $25\times$ the hardware time of \sys, which also includes graph and result transfers. End to end, the gap narrows to $3$ to $20\times$, because a 27\,$\mu$s Python-driven DMA call dominates small graphs on the FPGA. Against HyperX on the same board~\cite{hyperx2026}, batch-1 latency is $6$ to $42\times$ lower and throughput is $38$ to $317\times$ higher.

\begin{table*}[t]
\caption{Time and dynamic energy per graph (data of Fig.~\ref{fig:teaser}). Software runs Python end to end, and HyperX is quoted~\cite{hyperx2026} (--: not reported). \sys runs at batch one (mean latency) and at full throughput (hardware time). Energy is measured on the rail (Arm A53, \sys), package (CPU) or board (GPU), or estimated by tools (HyperX). All times are in $\mu$s and all energies in $\mu$J; lower is better. Shaded, bold entries denote \sys results.}
\label{tab:perf}
\centering
\footnotesize
\setlength{\tabcolsep}{4pt}
\begin{tabular}{@{}l r rrr r rr rrr r rr@{}}
\toprule
 & & \multicolumn{6}{c}{time per graph ($\mu$s)} & \multicolumn{6}{c}{dynamic energy per graph ($\mu$J)} \\
\cmidrule(lr){3-8}\cmidrule(l){9-14}
 & & \multicolumn{3}{c}{software, Python} & quoted & \multicolumn{2}{c}{\sys} & \multicolumn{3}{c}{software, Python} & quoted & \multicolumn{2}{c}{\sys} \\
\cmidrule(lr){3-5}\cmidrule(lr){6-6}\cmidrule(lr){7-8}\cmidrule(lr){9-11}\cmidrule(lr){12-12}\cmidrule(l){13-14}
Dataset & graphs & Arm A53 & CPU & GPU & HyperX & batch 1 & throughput & Arm A53 & CPU & GPU & HyperX & batch 1 & throughput \\
\midrule
MUTAG    & 188  & 14\,900 & 2\,410 & 66.9 & 1\,190 & \vesgacell{28} & \vesgacell{3.76} & 3\,550 & 25\,700 & 20\,400 & 970 & \vesgacell{8.7} & \vesgacell{5.3} \\
PTC\_FM  & 349  & 7\,980 & 1\,300 & 39.7 & -- & \vesgacell{22} & \vesgacell{2.93} & 1\,560 & 16\,300 & 11\,700 & -- & \vesgacell{4.3} & \vesgacell{3.5} \\
ENZYMES  & 600  & 29\,600 & 5\,300 & 119 & 450 & \vesgacell{42} & \vesgacell{5.84} & 5\,930 & 60\,300 & 44\,700 & 320 & \vesgacell{11.4} & \vesgacell{10.7} \\
PROTEINS & 1113 & 35\,200 & 6\,290 & 139 & -- & \vesgacell{46} & \vesgacell{6.46} & 7\,520 & 73\,100 & 51\,700 & -- & \vesgacell{9.4} & \vesgacell{12.5} \\
NCI1     & 4110 & 24\,400 & 3\,870 & 85.1 & 610 & \vesgacell{33} & \vesgacell{4.30} & 6\,010 & 44\,800 & 30\,600 & 480 & \vesgacell{8.4} & \vesgacell{8.0} \\
DD       & 1178 & 266\,000 & 51\,700 & 1\,100 & 1\,650 & \vesgacell{257} & \vesgacell{43.0} & 54\,500 & 627\,000 & 430\,000 & 1\,330 & \vesgacell{96.3} & \vesgacell{95.9} \\
\bottomrule
\end{tabular}
\end{table*}

\begin{figure}[t]
  \centering
  \includegraphics[width=\columnwidth]{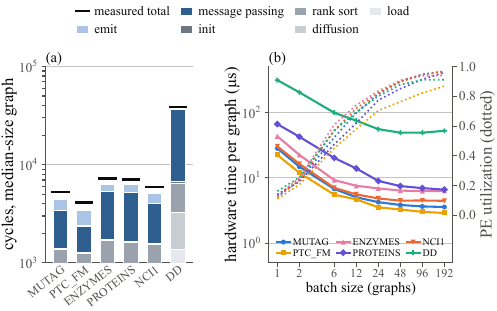}
  \caption{(a) Model cycles per stage (Eq.~\eqref{eq:cost}) for the median graph of each dataset, with the measured batch-1 latency (bar). (b) Hardware time per graph and PE utilization versus batch size.}
  \label{fig:latency}
\end{figure}

The latency follows a simple cycle model. A whole-graph record with $P=\sum_v\max\big(1,\lceil\deg^-(v)/2\rceil\big)$ edge pairs occupies a PE for approximately
\begin{align}
T \approx{}& \underbrace{(9+n+E)}_{\text{load}} + \underbrace{(K{-}1)(P{+}16)}_{\text{diffusion}} + \underbrace{(9n+1024)}_{\text{rank sort}} \nonumber\\
&+ \underbrace{(1+16L)(P{+}16)}_{\text{message passing}} + \underbrace{1024}_{\text{emit}}
\label{eq:cost}
\end{align}

cycles, where the per-pass constant accounts for pipeline drain. For a MUTAG graph, the model is within 3\,\% of the measured time per graph and PE. Fig.~\ref{fig:latency}a breaks the cycles down by stage. Message passing dominates on DD, while on the small graphs the rank sort and the fixed emission take a large share and set a floor of about 21\,$\mu$s. Transfer, write-out and drain add nearly constant overhead, and a linear fit over the 7187 whole-graph records with $K=4$ and $L=3$ gives 20\,$\mu$s plus 0.13\,$\mu$s per node and 0.14\,$\mu$s per edge ($R^2=0.9999$). As the batch grows, PE utilization rises above 0.9 at 48 graphs and the hardware time per graph converges to the throughput value (Fig.~\ref{fig:latency}b).

\subsection{Energy}
\label{sec:eval:energy}
All reported energies are dynamic (Sec.~\ref{sec:method}). As Table~\ref{tab:perf} shows, \sys needs a few microjoules per graph on the small benchmarks and about 96\,$\mu$J on DD. At batch one, only one PE is active, while the five idle PEs contribute negligible dynamic power. Activating all six PEs increases throughput and dynamic power together without reducing the work per graph. Energy per graph therefore remains similar, with batch-one energy at most about $1.6\times$ the throughput value. On the same rail, \sys uses $450$ to $750\times$ less energy than the embedded CPU. For the CPU and GPU comparisons, we use \sys's board-level energy, which includes the processing system, memory and regulators. This is 6.9 to 16.9\,$\mu$J per graph on the small benchmarks and 109\,$\mu$J on DD, still $2\,400$ to $5\,800\times$ below CPU package energy and $1\,700$ to $3\,900\times$ below GPU board energy. HyperX reports tool estimates rather than measured energy. Comparing both designs on that basis, \sys's estimated batch-one energy is $1.4$ to $8\times$ lower.

\subsection{Dimensionality and Large Graphs}
\label{sec:eval:scale}
\begin{figure}[t]
  \centering
  \includegraphics[width=\columnwidth]{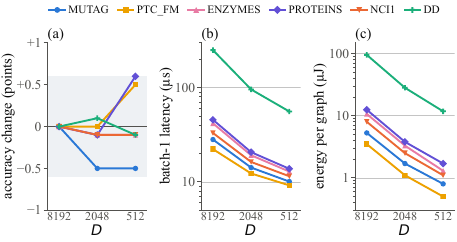}
  \caption{Measured builds at three widths $D$ (seed 0, 10 folds). (a) Accuracy change from $D=8192$ (shaded: $\pm0.6$ points). (b) Batch-1 latency. (c) Rail-dynamic energy per graph at full throughput.}
  \label{fig:dsweep}
\end{figure}

The co-designed kernel retains the robustness of \vsg to hypervector width. In the model, accuracy at every width from $D=8192$ down to 128 stays within 1.1 points of the $D=8192$ baseline. Because $D$ is a build-time parameter that does not affect the memory system, we also built and measured $D=2048$ and $D=512$ (Fig.~\ref{fig:dsweep}). Both are bit-exact with the model and within 0.6 points of $D=8192$, and both reduce latency and energy per graph, with energy falling by up to $8\times$ at $D=512$. Time falls by less than the $4\times$ and $16\times$ reduction in work, because the fixed front end is amortized over fewer slices. Only block RAM use shrinks, since area scales with $W$ and the PE count rather than with $D$. Exact splitting divides the two DD graphs of 4152 and 5748 nodes into four parts with 6 to 9\,\% halo overhead, and the complete DD set costs 43.0\,$\mu$s per graph, against 41.0\,$\mu$s without them, with bit-identical embeddings.

\subsection{Beyond Nominal Operation}
\label{sec:eval:robust}
\begin{figure}[t]
  \centering
  \includegraphics[width=\columnwidth]{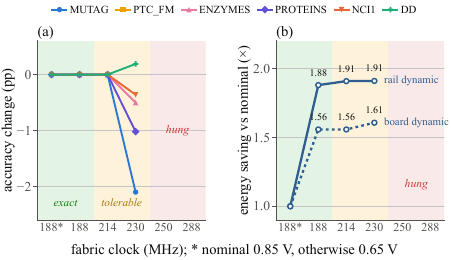}
  \caption{Measured operation beyond the sign-off point. (a) Accuracy change per dataset. (b) Saving in dynamic energy per graph (rail and board) relative to nominal, mean over the datasets. PTC\_FM hangs at 230\,MHz and every dataset from 250\,MHz.}
  \label{fig:robust}
\end{figure}

Undervolting in HDC hardware remains largely unexplored, with recent
studies primarily focusing on watermarking reliability
\cite{poursiami2026hdmark,poursiami2026intellectual}.
Here, we investigate its impact on computational correctness,
classification accuracy, and energy efficiency.

Because each lane uses majority voting and the readout counts bits, sporadic bit errors may be tolerated. In a fault-injection model, random flips of hypervector reads, code bits or lane results at rates up to $10^{-3}$ change accuracy by at most 0.4 points, and at $10^{-1}$ by at most 1.4 points. On the board, we lowered VCCINT from 0.85 to 0.65\,V, the regulator floor, and raised the fabric clock from 187.5 to 287.5\,MHz. At each point we compared every embedding with the nominal one, re-ran the classifier, and measured latency and power (Fig.~\ref{fig:robust}). A point is \emph{exact} if all rows are bit-exact, \emph{tolerable} if some rows are wrong but accuracy stays within one fold standard deviation, and \emph{hung} otherwise.

Undervolting remains bit-exact. At 0.65\,V and the design clock, all 7538 graphs stay bit-exact and the rail-dynamic energy per graph halves (Fig.~\ref{fig:robust}b). The first errors appear at 214\,MHz, only in the rows of one PE and in one 64-lane group, in the block that the timing report names as critical. Accuracy is unchanged on all datasets, so this point is tolerable at 12\,\% lower latency. At 230\,MHz the same PE corrupts most lanes of its rows, yet accuracy stays within one fold standard deviation on five datasets (Fig.~\ref{fig:robust}a), while PTC\_FM hangs in the dispatcher handshake. At 250\,MHz a second PE fails and the DMA path wedges, and only reprogramming the fabric recovers it. As the model predicted, the datapath degrades gracefully but the control path does not, and the clock rather than the voltage sets the frontier. For deployment, 0.65\,V at 187.5\,MHz halves the dynamic energy without observed errors, and 214\,MHz adds 12\,\% speed for an error that the classifier absorbs.
\section{Conclusion}
\label{sec:conclusion}
Co-designing the \vsg kernel and its accelerator turned a model that takes milliseconds per graph on a CPU into a microsecond-scale, microjoule-scale workload on an embedded FPGA, without a measurable accuracy cost. Most gains come from three choices: a one-bit state per node and lane, a computed code book, and a slice schedule that loads the next slice for free. Because the arithmetic is a majority vote, the accelerator tolerates operation beyond its sign-off point, and we measured its failure modes. These results show that brain-inspired HDC graph learning maps efficiently onto simple hardware and suggest ASIC and in-memory realizations as future work.

\bibliographystyle{IEEEtran}
\bibliography{IEEEabrv,refs}

\end{document}